\documentclass[aps, twocolumn, groupedaddress,superscriptaddress, showpacs,pre,showkeys]{revtex4-1}
\usepackage{graphicx}
\usepackage[title]{appendix}
\usepackage{array}
\usepackage{setspace,transparent}
\usepackage{color}
\usepackage{amsmath}
\usepackage{chngcntr}
\usepackage{fontenc,subcaption,ragged2e,amsmath}
\usepackage{mathtools,amssymb,amssymb,nicefrac,tikz}
\usepackage{soul}
\usepackage[font=small,labelfont=bf]{caption}
\usepackage[utf8]{inputenc}

\providecommand{\keywords}[1]{\textbf{\textit{Index terms---}} #1}

\DeclareCaptionJustification{justified}{\justifying}
\newcommand{\RN}[1]{%
	\textup{\uppercase\expandafter{\romannumeral#1}}%
}
\DeclareMathAlphabet{\mathpzc}{OT1}{pzc}{m}{it}
\DeclareCaptionJustification{justified}{\justifying}
\begin{document}
\title{Interdependent transport via percolation backbones in spatial networks}
\author{Bnaya Gross}
\email{bnaya.gross@gmail.com}
\affiliation{Department of Physics, Bar-Ilan University, Ramat-Gan 52900, Israel}
\author{Ivan Bonamassa}
\affiliation{Department of Physics, Bar-Ilan University, Ramat-Gan 52900, Israel}
\author{Shlomo Havlin}
\affiliation{Department of Physics, Bar-Ilan University, Ramat-Gan 52900, Israel}

\date{\today}

\begin{abstract}
	The functionality of nodes in a network is often described by the structural feature of belonging to the giant component. However, when dealing with problems like transport, a more appropriate functionality criterion is for a node to belong to the network's  backbone, where the flow of information and of other physical quantities (such as current) occurs. Here we study percolation in a model of interdependent resistor networks and show the effect of spatiality on their coupled functioning. We do this on a realistic model of spatial networks, featuring a Poisson distribution of link-lengths. We find that interdependent resistor networks are significantly more vulnerable than their percolation-based counterparts, featuring first-order phase transitions at link-lengths where the mutual giant component still emerges continuously.  We explain this apparent contradiction by tracing the origin of the increased vulnerability of interdependent transport to the crucial role played by the dandling ends. 
	Moreover, we interpret these differences by considering an heterogeneous $k$-core percolation process which enables to define a one-parameter family of functionality criteria whose constraints become more and more stringent. Our results highlight the importance that different definitions of nodes functionality have on the collective properties of coupled processes, and provide better understanding of the problem of interdependent transport in many real-world networks.\\
	\\
	\textit{This work is dedicated to the late Prof. Dietrich Stauffer from whom we learned a lot about percolation.}
\end{abstract}


\maketitle
\section{Introduction}
Throughout the last decades, network science has provided important tools to study complex systems such as the brain \cite{moretti2013griffiths_critical_brain_hypothesis,sporns2004organization}, climate networks \cite{jingfan2017network_climate,donges2009complex}, protein interactions \cite{kovacs2019network,de2015structural} and finance \cite{stauffer1999self,onnela2003dynamics,stauffer2000sharp}, offering a powerful framework for exploring their collective phenomena \cite{dorogovtsev2008critical}. The ability to simplify a complex system to its basic ingredients and still observing the general phenomenon occurring in it is, perhaps, one of the main reasons for the rise of network science in recent years. \par
A prominent tool commonly used in the analysis of the structure and function of many real-world networks  is percolation theory \cite{stauffer2018introduction,bunde2012fractals,pandey1983confirmation}. During this process, a fraction $1-p$ of nodes or edges are randomly removed and certain quantities of interest such as the giant component (GC), the correlation length, or the susceptibility, are then measured. For sufficiently large values of $p$, a giant component spanning the entire network exists, enabling the communication between nodes belonging to it, and at a critical threshold $p_c$ it dismantles into a collection of small clusters.\par
\begin{figure}
	\centering
	\begin{tikzpicture}[      
	every node/.style={anchor=north east,inner sep=0pt},
	x=1mm, y=1mm,
	]   
	\node (fig3) at (0,0)
	{\includegraphics[scale=0.5]{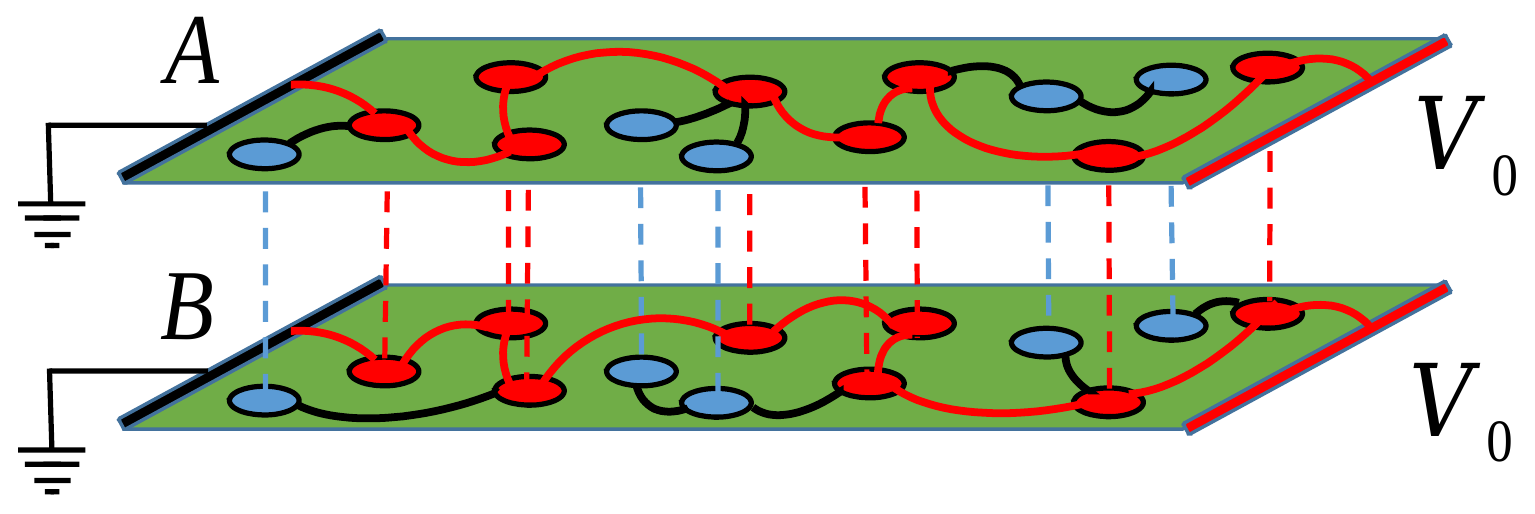}};
	\end{tikzpicture}
	\caption{\textbf{Illustration of the interdependent resistor model}. Transport of currents in two networks of spatial resistors, A and B, are mutually coupled via dependency links (dashed lines). Each layer is constructed with links of the same characteristic length $\zeta$ as described in Eq. \eqref{eq:zeta_prob} and the same average degree $z$, though their local wiring features are generally different. The backbone in each layer consists of the red nodes connected via the red connectivity links which conduct current between the network's boundaries. The blue nodes do not conduct current (dead ends) and thus belongs only to the giant component but not to the backbone. A node will fail if it is not part of the backbone of its network or if its dependent node in the other layer fails.}
	\label{fig:Illustration}	
\end{figure}
The functionality of the network is usually described by adopting as a proxy the relative size of the GC, $P_{\infty}$, and nodes that disconnect from it are isolated and considered as non-functional. However, when transport processes like e.g.\ current flow in resistor networks \cite{kirkpatrick1971classical,derrida1983transfer,kirkpatrick1973percolation} are considered, a more appropriate criterion for the nodes' functionality has to be introduced. For resistor networks, such condition can be identified in the requirement that a node belongs to the relative size of the network's \textit{backbone} \cite{bunde2012fractals,kirkpatrick1973percolation}, $B_{\infty}$, which contains only conduct-current nodes, i.e. no dead-ends (see Fig. \ref{fig:Illustration}). \par
 
The importance of these differences in the definition of the nodes' functionality becomes more significant when considering multilayer networks \cite{bianconi2018multilayer,de2013mathematical} and, in particular, interdependent networks \cite{buldyrev2010catastrophic,stippinger2014enhancing,gao2012interdependentnetworks,baxter2012avalanche,gross2020interconnections,radicchi2015percolation}.  
In such cases, the failing of a node in one network can cause further damage in other one, which can in its turn trigger a cascade of failures resulting in abrupt collapses signaled by first-order structural transitions. Since failed nodes are the ones spreading the damage from one network to the other, the precise definition of the functionality criterion becomes a crucial ingredient in understanding the vulnerability and characterizing the functional regimes of interdependent systems. 
\par
In this paper, we study percolation on a model of two interdependent resistor networks with conductivity-based states, so that global transport is attained only if a mutual backbone exists.  Motivated by recent evidence on transport networks \cite{danziger2016effect,halu2014emergence} and in the connectome's structure of mammals \cite{bullmore2012economy,markov2014weighted,ercsey2013predictive,horvat2016spatial}, we consider here the realistic case of spatially embedded networks with a tunable characteristic link length \cite{danziger2016effect,gross2017multi,vaknin2017spreading,bonamassa2019critical}. We find that, in contrast to a single network where both the GC and backbone have the same critical threshold, in interdependent networks the critical thresholds signalling the collapse of the giant components are different. In particular, we show that the critical threshold for the backbone is much higher compared with its percolation-based analogue, hinting at the extreme vulnerability of interdependent transport in spatially embedded networks. In addition, while the transition changes from second to first order as the interaction range (link-length) increases for both the GC and backbone, the backbone transition becomes first order in a much shorter interaction range compared to the GC. Furthermore, using heterogeneous k-core percolation \cite{cellai2011tricritical,panduranga2017generalized,baxter2011heterogeneous}, we are able to explain the reason for the shorter interaction range required to trigger first-order transitions in interdependent resistor networks. We show that as the criteria for node functionality gets more strict, the damage can spread in the whole system with shorter interaction range. \par
We stress that the cost function here considered is more realistic than those of previous studies \cite{danziger2015interdependent}, and motivated by data-driven evidence reported in transport systems \cite{danziger2016effect,halu2014emergence} and in brain networks \cite{bullmore2012economy,markov2014weighted,ercsey2013predictive,horvat2016spatial}. 
In this respect, our results provide additional insights to the properties of coupled transport processes in spatial infrastructures \cite{morris2012transport,morris2013interdependent,morris2014spatial}, offering a simple and realistic framework to investigate their robustness. 
 \begin{figure}
 	\centering
 	\begin{tikzpicture}[      
 	every node/.style={anchor=north east,inner sep=0pt},
 	x=1mm, y=1mm,
 	]   
 	\node (fig3) at (0,0)
 	{\includegraphics[scale=0.5]{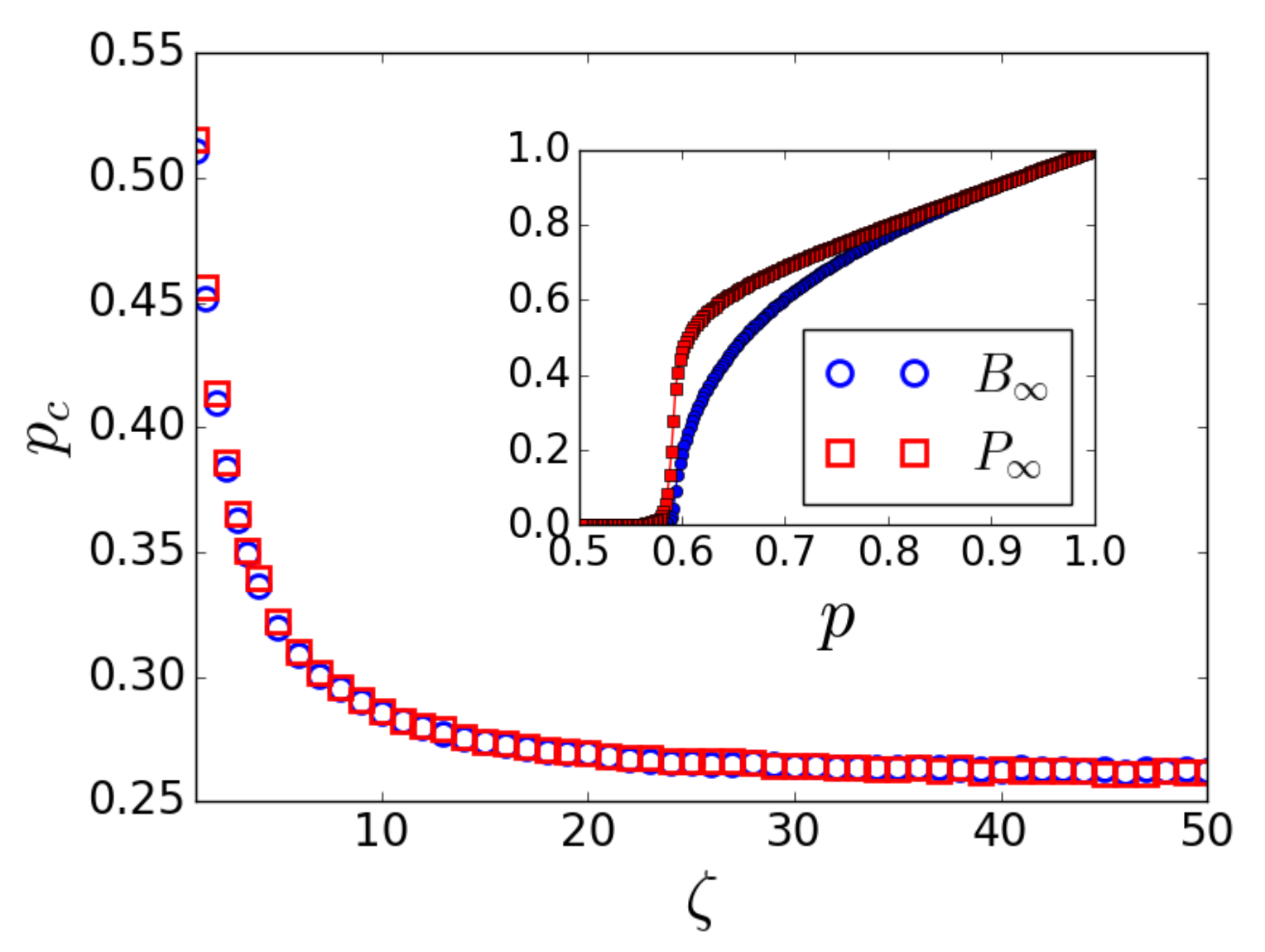}};
 	\end{tikzpicture}
 	\caption{\textbf{Percolation and conductivity thresholds in a single spatial network.} Both $P_{\infty}$ and $B_{\infty}$ have the same percolation threshold $p_c$ for any value of $\zeta$. Adopting $z=4$, one has in the limit of $\zeta \ll 1$ only short links and a 2D lattice-like structure is created with the known $p^{2D}_c \simeq 0.5926$ \cite{stauffer2018introduction,bunde2012fractals}. In the other limit of $\zeta \to \infty$ any pair of nodes can be connected with the same probability similar to an ER network with $p^{ER}_c = 1/z$. The inset shows the size of the giant component (GC) and the backbone of a single 2D lattice. Notice that the GC contains also nodes that do not conduct current (dead ends, see Fig. \ref{fig:Illustration}) and thus the backbone is a sub-set of the GC, while the transition occurs at the same percolation threshold. Here and throughout the paper, simulation results are obtained for networks of size $N=10^6$.}
 	\label{fig:giant_backbone_lattice}
 \end{figure}
\section{The model}
We model interdependent transport by means of two spatial networks, A and B, as depicted in  Fig.~\ref{fig:Illustration}. The nodes in each layer are placed on a 2-dimensional grid of size $N = L \times L$, where $L$ is the grid length, on the positions $(x,y)$ where $x,y \in [0,L-1]$ are integers numbers. The connectivity links in each network are then assigned by picking randomly a node $i$ and connecting it with a random node $j$ at Euclidean distance $\mathit{d}_{ij}$ drawn from an exponential distribution
\begin{equation}
    \mathcal{P}_{ij}(\mathit{d}_{ij}) \propto \exp(-\mathit{d}_{ij}/\zeta) .
    \label{eq:zeta_prob}
\end{equation}
Here, $\zeta$ represents the characteristic link-length of the network and plays the role of a tunable parameter controlling the influence of spatiality on the range of interactions. This picking process repeats until a given average degree $z$ is reached. As discussed in earlier works by some of us \cite{bonamassa2019critical,danziger2020faster}, the structure of the network significantly depends on the characteristic link-length $\zeta$: while small values of $\zeta$ produce strongly space-dependent networks, large values of $\zeta$ (order $\mathcal{O}(L)$) produce networks with weak space-dependence which can be analyzed via mean-field approaches \cite{danziger2016effect,gross2017multi,vaknin2017spreading,bonamassa2019critical}. The two networks depend on each other through dependency links between nodes placed in the same geometrical position in both networks (see Fig. \ref{fig:Illustration}). Therefore, if the node $(x,y)$ fails in layer A, then also its ``replicated'' node $(x,y)$ in layer B will fail.  Let us stress that the neighbourhoods of superposed nodes in the two layers are  generally different, since each layer is a different instance of the same statistical ensemble of spatially embedded networks. \par
We study percolation on our interdependent model with conductivity-based functionality by removing non-conducting nodes that do not belong to the percolation \textit{backbone} (the dandling ends) of each layer, which we measure by searching for the networks' largest bi-components \cite{grassberger1999conductivity}. The process is initiated by removing a fraction $1-p$ of nodes from network A. This removal may disconnect some nodes from the backbone of network A causing their dependent nodes in network B to be removed as well. The removal of nodes in network B may disconnect more nodes from the backbone of network B which, in their turn, make their dependent nodes in network A to fail, hence propagating the damage. This repeating \textit{cascade of failures} describes the dynamic behavior of the system and it is an intrinsic property of interdependent networks and their stability. Once the cascading process stops, the remaining active nodes in the whole system form the \textit{mutual backbone} (MB). Similarly, the remaining active nodes after the cascading process with only percolation-based functionality form the so-called mutual giant component (MGC) of the system.  Notice that, although both the MB and the MGC are respectively, subsets of the backbone and giant component in their isolated counterparts, they are measured respectively by means of the very same observables, namely $B_{\infty}$ and $P_\infty$. \par

\section{The effects of functionality on the phase transition}
To understand the significant difference between percolation-based functionality and conductivity-based functionality in interdependent networks, let us first consider the case of a single isolated layer. Percolation in a single network yields a continuous structural transition at the same position for both the GC and the backbone (see Fig.~\ref{fig:giant_backbone_lattice}). The reason is that a path from one side of the network to the other exists even if non-conducting nodes (dandling ends) are removed \cite{stauffer2018introduction,bunde2012fractals} and thus their removal only affect the magnitude of the order parameter without changing the transition threshold (see Fig.~\ref{fig:giant_backbone_lattice}, inset). In the limit of $\zeta \ll 1$, only short link-lengths are allowed and a 2D lattice-like structure is created, with $p^{2D}_c \simeq 0.5926$ \cite{stauffer2018introduction,bunde2012fractals} for both the GC and the backbone. In the other limit, i.e.\ $\zeta \to \infty$, any pair of nodes can be connected with the same probability similar to an ER network, leading therefore to the percolation threshold $p^{ER}_c = 1/z$. Notice that $p_c$ rapidly converges towards $p^{ER}_c$ (see Fig.~\ref{fig:giant_backbone_lattice}), resulting in a 2D-to-random crossover with surprising features, whose details were extensively addressed in Ref.~\cite{bonamassa2019critical}. \par
\begin{figure}
    \centering
    \begin{tikzpicture}[      
	every node/.style={anchor=north east,inner sep=0pt},
	x=1mm, y=1mm,
	]   
	\node (fig3) at (0,0)
	{\includegraphics[scale=0.2]{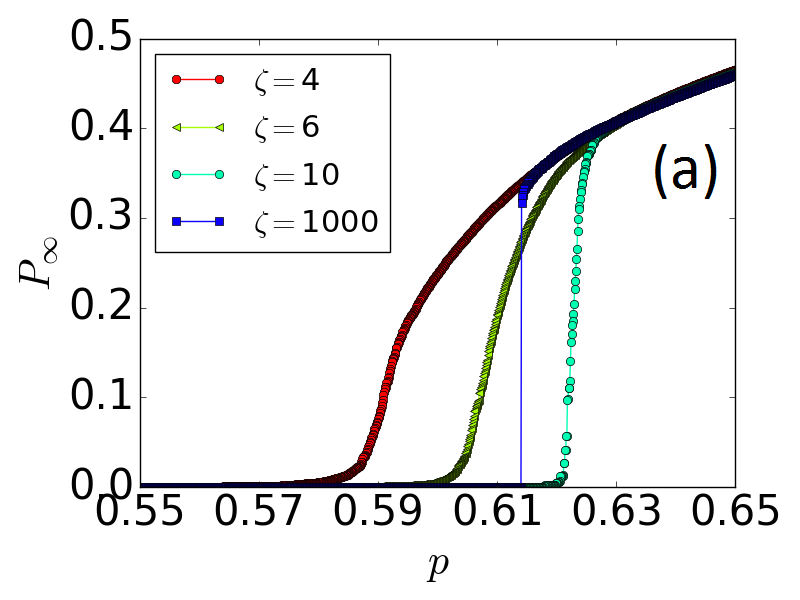}};
	\node (fig3) at (40,-1)
	{\includegraphics[scale=0.2]{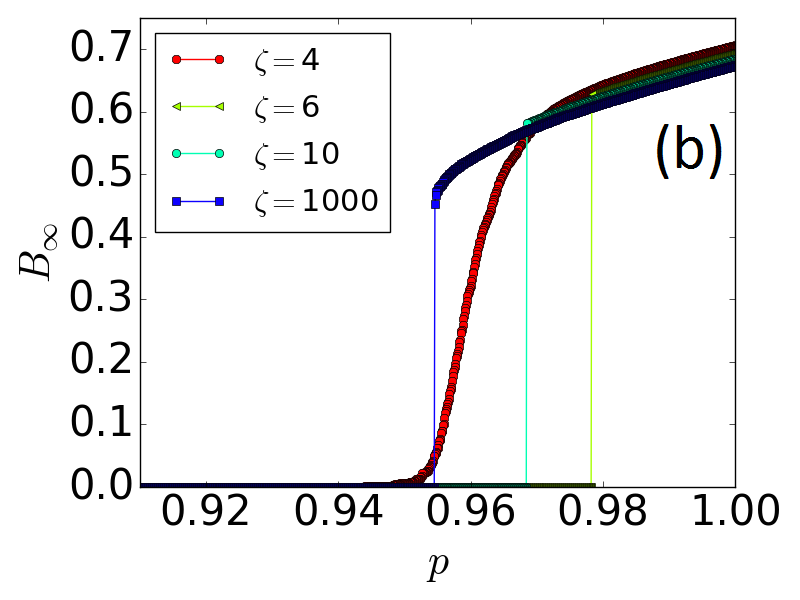}};
	\end{tikzpicture}
	\caption{\textbf{Interdependent percolation and conductivity transitions.} The relative size of the (a) MGC, $P_{\infty}$, and the (b) MB, $B_{\infty}$, as a function of $p$ for several values of $\zeta$ are shown. For small values of $\zeta$ the transition is continuous for both the MGC and the MB. However, as $\zeta$ exceeds a critical interaction length, $\zeta_c$, the transition becomes first-order. Notice that $\zeta_c$ of the MGC is larger compared to that of the MB.}
	\label{fig:giant_backbone}	
\end{figure}
\begin{figure}
    \centering
    \begin{tikzpicture}[      
	every node/.style={anchor=north east,inner sep=0pt},
	x=1mm, y=1mm,
	]   
	\node (fig3) at (0,0)
	{\includegraphics[scale=0.47]{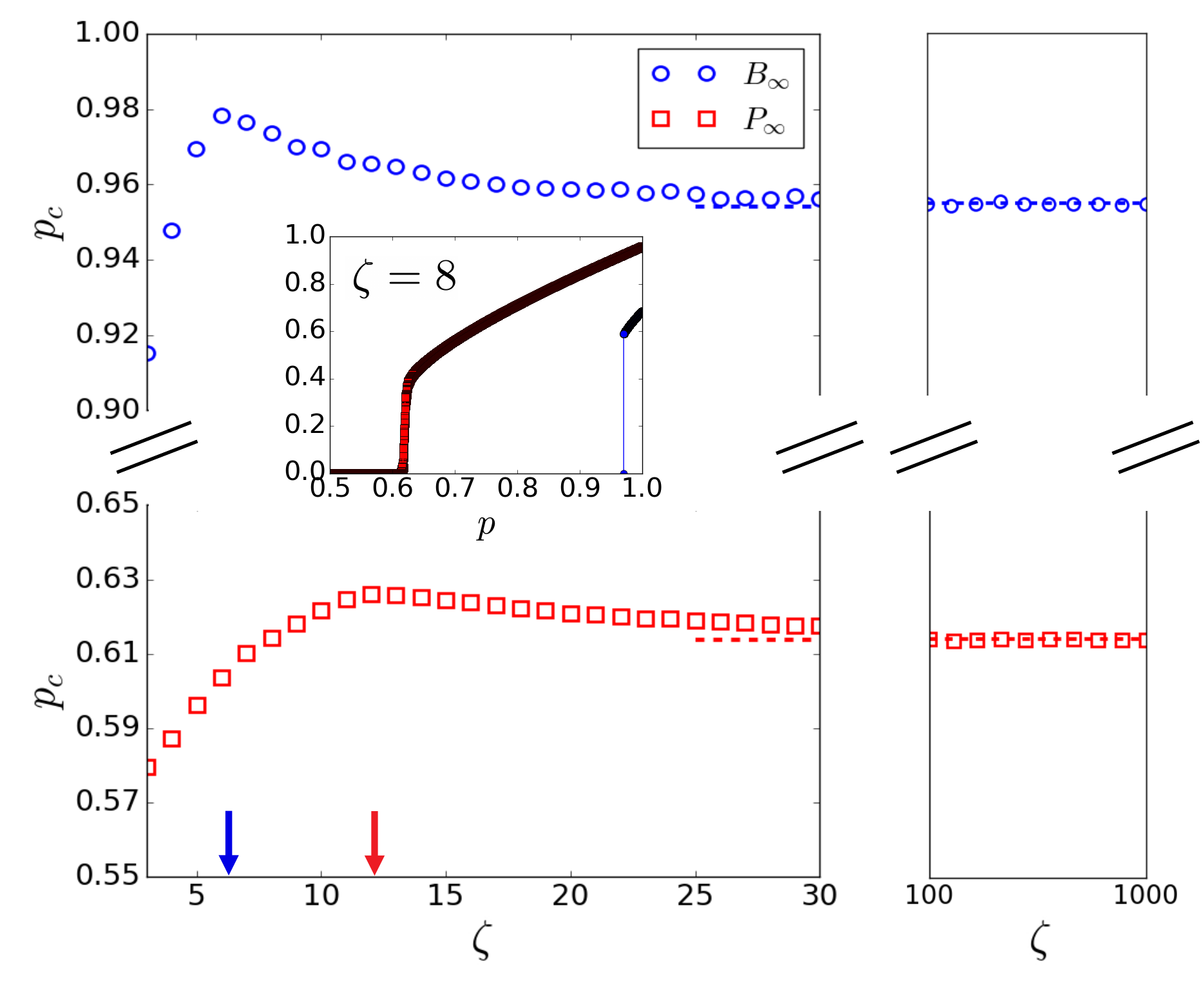}};

	\end{tikzpicture}
	\caption{\textbf{Critical thresholds.} Phase diagram showing the critical thresholds $p_c$ for the MGC (red) and the MB (blue) change with increasing values of $\zeta$. For $\zeta < \zeta_c$ both transitions are continuous and $p_c$ increases close to linearly with $\zeta$, reaching a peak at $\zeta_c$. For $\zeta > \zeta_c$, the transitions are instead first-order and both $p_c$ slowly decrease, converging to the mean-field value. In the limit of random interactions, i.e.  $\zeta \to \infty$, $p_c \to 2.4554/z$ for the MGC (red dashed line) and $p_c \approx 3.8166/z$ (blue dashed line) for the MB for $z=4$. Notice that the value of $\zeta_c$ of the MB is smaller compared to that of the MGC (approximately 6 for  MB and 12 for MGC), unveiling a region where, even if the MGC undergoes a continuous phase transition, the MB collapses abruptly. The inset demonstrates this phenomenon for $\zeta = 8$.}
	\label{fig:giant_backbone_pc}	
\end{figure}

Interdependent networks experience completely different phenomena compared to a single network. For the case of the percolation-based functionality \cite{danziger2016effect}, in the limit $\zeta \to \infty$ (two interdependent ER networks) the percolation phase transition becomes first-order as shown in Fig.~\ref{fig:giant_backbone}a, and it can be analytically solved, resulting in the critical threshold $p_c \simeq 2.4554/z$ \cite{buldyrev2010catastrophic} (see Fig.~\ref{fig:giant_backbone_pc}). Moreover, a tricritical characteristic length $\zeta_c \simeq 12$ exists above which a local damage will propagate at distances sufficiently large (i.e. larger than the radius of a critical droplet \cite{unger1984nucleation,congilio1980clusters,heermann1983nucleation}) igniting a percolative nucleation process \cite{danziger2016effect} that leads to a first-order phase transition.  
In contrast, for $\zeta < \zeta_c$ local failures generally remain confined, leading to continuous phase transitions whose cluster statistics is strongly influenced by finite-size effects \cite{li2012cascading}. \par
The case of conductivity-based interdependence, discloses important differences compared to its percolation-based analogue. The first difference can be identified in the transition point, which is not in the same position as can be seen in Fig.~\ref{fig:giant_backbone}. This is in marked contrast to a single network case where the transition point is in the same position (Fig.~\ref{fig:giant_backbone_lattice}). The reason for this difference can be understood in the effect of the dangling ends. For a single network, joining the network's boundaries exists even after the removal of the dangling ends, thus, their removal does not affect the transition threshold. However, once dependency links between networks are set, the removal of the dandling ends in one network can lead to failure of nodes belonging to the backbone of the other network, a genuine multilayer effect that finds no analogy in the isolated case. This removal leads to a much stronger cascade of failures in the system compare to percolation-based functionality and breaks the path joining the network's boundaries. These cascades lead to the separation of the transition of percolation-based functionality and conductivity-based functionality observed even at small interaction ranges and it further explains the origin underlying the extreme vulnerability of the MB. Another important difference between the MGC and the MB is the transition behaviour for different values of $\zeta$. Similarly to the MGC, the MB undergoes a first-order transition for $\zeta \to \infty$ as shown in Fig.~\ref{fig:giant_backbone}b whose features can be solved analytically (see Appendix), resulting in the threshold $p_c \simeq 3.8166/z$ as shown in Fig.~\ref{fig:giant_backbone_pc}. However, the value of $\zeta_c$ is much smaller: while for the MGC, $\zeta_c \simeq 12$, for the MB, $\zeta_c \simeq 6$, as shown in Fig.~\ref{fig:giant_backbone_pc}. \par
\section{Tricritical points in the characteristic range of interactions}
In order to better understand the drastic decrease of the tricritical interaction range $\zeta_c$ for conductivity-based functionality systems, we here examine heterogeneous $k$-core percolation on our interdependent spatial network model. Let us recall that $k$-core percolation is an iterative process initiated by random removal of $1-p$ fraction of nodes followed by iterative removal of nodes with degree less than $k$ until only the $k$-core remains \cite{dorogovtsev2006k}. In \textit{heterogeneous} $k$-core percolation, the degree threshold is not the same for all the nodes \cite{cellai2011tricritical,panduranga2017generalized,baxter2011heterogeneous}, and it is assigned in a way such that an $r$ fraction of randomly chosen nodes has threshold $k_a$ and the remaining fraction $1-r$ has threshold $k_a + 1$. Thus, the average degree threshold is given by
\begin{equation}
	k = k_a(1-r) + r(k_a+1) .
	\label{eq:k_thresh}
\end{equation}
By continuously increasing $r$, we study the effect of node functionality on the system's phase transitions as it gets increasingly more stringent. Eq.~\eqref{eq:k_thresh}, in fact, allows to identify a one-parameter family of functionality criteria for each $r$ so that different levels of functionality constrains can be compared. \par
\begin{figure}
	\centering
	\begin{tikzpicture}[      
	every node/.style={anchor=north east,inner sep=0pt},
	x=1mm, y=1mm,
	]   
	\node (fig3) at (0,0)
	{\includegraphics[scale=0.37]{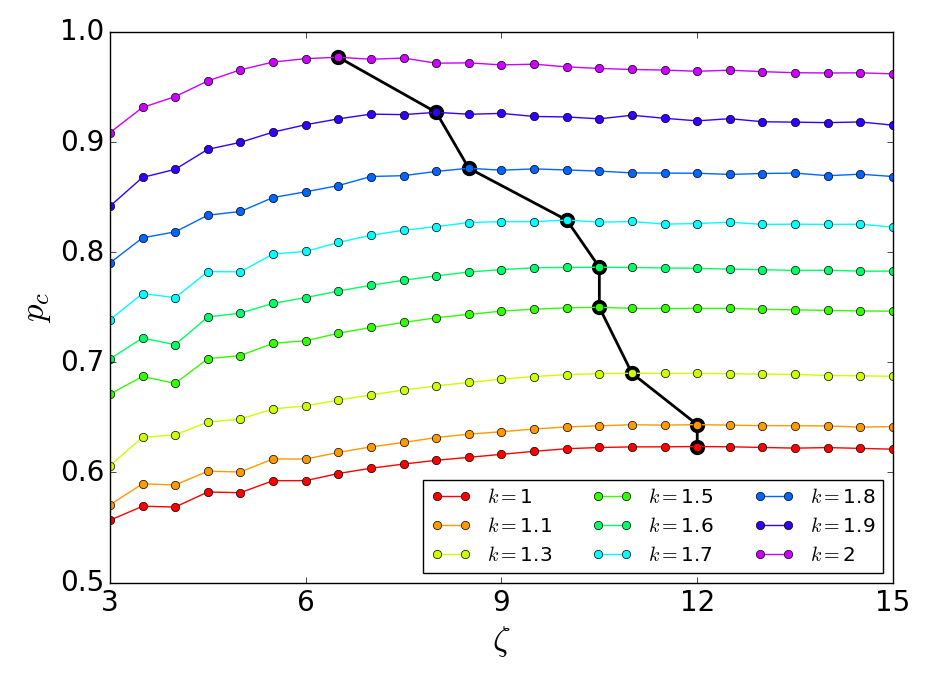}};
	
	\end{tikzpicture}
	\caption{\textbf{Phase diagram for interdependent $k$-core.} $p_c$ is measured as a function of $\zeta$ for different values of average degree threshold, $k$, as calculated from Eq.~\eqref{eq:k_thresh} with $k_a = 1$. As expected for $k = 1$ the case of MGC is recovered with $\zeta_c \simeq 12$. However, as $k$ increases and the nodes functionality criterion gets more strict, $\zeta_c$ decreases as shown by the black line. The case of $k = 2$ shows the same critical interaction range as the MB ($\zeta_c \approx 6$) even though 2-core percolation and the backbone are not exactly the same since node can have degree 2 but not be part of the backbone.}
	\label{fig:heter_kcore}	
\end{figure}

We start with the case, $k_a = 1$ and $r = 0$, which corresponds to the MGC, and increase $r$ to study how the tricritical interaction range $\zeta_c$ will change with $k$. The phase diagram in  Fig.~\ref{fig:heter_kcore} discloses the dependence of the percolation threshold as a function of $\zeta$ for different average degree thresholds. As expected, for $r = 0$ and $k=1$ we find $\zeta_c \simeq 12$, as in Fig.~\ref{fig:giant_backbone_pc} for the MGC. However, as the average degree threshold, $k$, increases, $\zeta_c$ decreases. This show that as the node functionality gets more strict, not only that the percolation threshold increases but the critical interaction range decreases. In other words, the MB has a much lower tricritical interaction range compared to the MGC , leading to a cascade of failures and abrupt collapses already at a relatively small range of interactions. In line with evidence raised by previous results in interdependent transport processes in spatial networks \cite{morris2012transport,morris2013interdependent,morris2014spatial}, our results highlight the dramatic fragility of infrastructures and transport systems. 
\section{Summary and Discussion}
In this work, we have studied the effect of spatiality on interdependent resistor networks emphasizing the differences between percolation-base functionality governed by the GC and conductivity-based functionality governed by the backbone. Our model makes a step forward towards a more realistic characterization of interdependent transport processes in real-world systems, thanks to the realistic spatial topology we have considered. We find that while in a single network the percolation transition is the same for both functionality criterion, once dependency links are formed between networks the transition thresholds are significantly different with higher vulnerability for the backbone. Moreover, both criteria have a tricritical interaction length above which the structural transitions are first-order and continuous below. We also find that the tricritical interaction length for the MB is shorter compared to that of the MGC, highlighting the extreme vulnerability of interdependent transport processes \cite{morris2012transport,morris2013interdependent,morris2014spatial}. We have explained this difference by adopting a model of interdependent heterogeneous $k$-core, showing that the tricritical interaction range decrease as the criterion for the nodes functionality gets more strict. \par
Our results highlight the crucial role played by the definition of node functionality which significantly affects its robustness against random failures, and offer new perspectives regarding the influence that precise definitions of nodes' functionality can have on their coupled collective phenomena. 
For example, a system of real interdependent networks might be characterized by percolation-based functionality in one layer and conductivity-based functionality in another, an outcome that would lead to critical features in between the two cases studied here. Moreover, in systems with even stricter node functionality criteria, e.g.\ governed by heterogeneous $k$-core with $k>2$ or involving more than 2 interacting layers, we expect that even more extreme critical properties will be found, with a larger increase in their vulnerability. It would therefore be of interest to investigate such cases where the addition of layers is accompanied by redundant interdependencies, so to compare how spatiality influences the increase of robustness with respect to the random case already addressed in the literature \cite{radicchi2017redundant}.
\section{Acknowledgements}
We thank the Israel Science Foundation, the Binational Israel-China Science Foundation Grant no.\ 3132/19, ONR, the BIU Center for Research in Applied Cryptography and Cyber Security, NSF-BSF Grant no.\ 2019740, and DTRA Grant no.\ HDTRA-1-19-1-0016 for financial support.
\numberwithin{equation}{subsection}
\numberwithin{figure}{subsection}
\section*{Appendix}
\subsection{Analytic solution of the limit $\zeta \to \infty$ for interdependent resistor networks}
The analytical solution can be developed using the generating function approach developed by Newman \cite{newman2001random}. The generating function for the degree distribution is 
\begin{equation}
G_0(u) = \sum\limits_{k} p_ku^k
\label{G0}
\end{equation}
and the generating function for the outgoing links is
\begin{equation}
G_1(u) = \frac{G^{'}_0(u)}{z} ,
\label{G1}
\end{equation}
\begin{figure}
	\centering
	\begin{tikzpicture}[      
	every node/.style={anchor=north east,inner sep=0pt},
	x=1mm, y=1mm,
	]   
	\node (fig1) at (-60,0)
	{\includegraphics[scale=0.85]{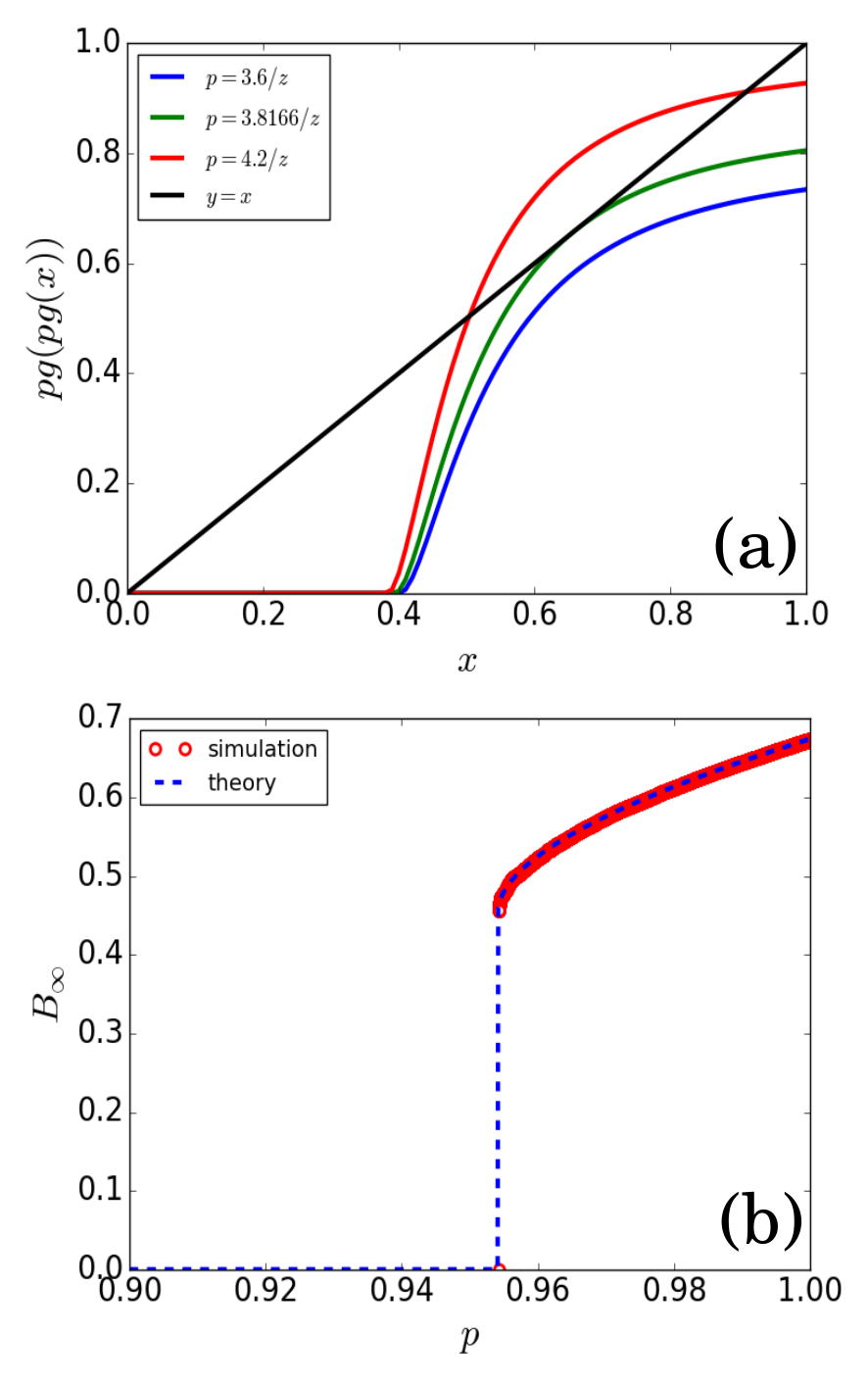}};
	\end{tikzpicture}
	\caption{\textbf{(a) Graphical solution of Eq.~\eqref{eq:x_self-consistent}}. The solution $x = 0$ always exist for any value of $p$. However, for $p_c = z_c/z$ with $z_c \approx 3.8166$ a non-zero solution appears whose value increases continuously with $z$. \textbf{(b) First-order phase transition.} When the solution of Eq.~\eqref{eq:x_self-consistent} is inserted into Eq~\eqref{eq:backbone_thoery}, one finds two regimes. Inactive state ($B_{\infty} = 0$, i.e $x = 0$), and an active state (i.e. $B_{\infty} > 0$ or, equivalently, $x\neq0$).}
	\label{fig:theory_simulations}	
\end{figure}
where $z$ is the average degree. \par
In percolation, the backbone is equivalent to the largest bi-component \cite{grassberger1999conductivity}. Thus, we can adopt the formalism developed in Ref. \cite{newman2008bicomponents} to calculate the mutual bi-component in random graphs. 
For a \textit{single network}, the probability $u(p)$ that a link does not lead to a node in the giant bi-component, after removal of $1-p$ fraction of nodes is \cite{newman2008bicomponents}:
\begin{equation}
u(p) = 1 - p + pG_1(u(p)),
\end{equation}
and the fraction of nodes belonging to the giant bi-component is:
\begin{equation}
\begin{split}
B_{\infty} & = pg(p) \\
& = p[1 - G_0(u(p)) - (1 - u(p))zG_1(u(p))] .
\end{split}
\end{equation}
For the case of ER networks, $p_k = \frac{e^{-z}z^{k}}{k!}$ and thus,
\begin{equation}
G_0(u) = G_1(u) = e^{-z(1-u)},
\end{equation}
with
\begin{align} 
u(p) &= 1 - p + pe^{-z(1-u(p))} \\
g(p) &= 1 - (1+(1-u(p))z)e^{-z(1-u(p))} .
\end{align}
\par
For the case of two interdependent networks, the size of the mutual bi-component is hence given by \cite{buldyrev2010catastrophic}:
\begin{equation}
B_{\infty} = xg(x),
\label{eq:backbone_thoery}
\end{equation}
where $x$ is the solution to the self-consistent equation
\begin{equation}
x = pg[pg(x)] .
\label{eq:x_self-consistent}
\end{equation}
Eq.~\eqref{eq:x_self-consistent} can be graphically solved, as shown in Fig.~\ref{fig:theory_simulations}a. The solution $x = 0$ always exist for any value of $p$. However, for $p_c = z_c/z$ with $z_c \approx 3.8166$, a non-zero solution appears which increases continuously with $z$. When inserted into Eq~\eqref{eq:backbone_thoery}, the graphical solution of Eq.~\eqref{eq:x_self-consistent} identifies two regimes: inactive state with $B_{\infty} = 0$ corresponding to $x = 0$, and an active state with $B_{\infty} > 0$ corresponding to the non-zero solution. The transition between these two states is abrupt, characterizing a (random) first-order phase transition with $p_c = z_c/z$ as shown in Fig.~\ref{fig:theory_simulations}b.
\bibliographystyle{unsrt}

\begin{thebibliography}{10}
	
	\bibitem{moretti2013griffiths_critical_brain_hypothesis}
	Paolo Moretti and Miguel~A Mu{\~n}oz.
	\newblock Griffiths phases and the stretching of criticality in brain networks.
	\newblock {\em Nature communications}, 4:2521, 2013.
	
	\bibitem{sporns2004organization}
	Olaf Sporns et~al.
	\newblock Organization, development and function of complex brain networks.
	\newblock {\em Trends in cognitive sciences}, 8(9):418--425, 2004.
	
	\bibitem{jingfan2017network_climate}
	Jingfang Fan et~al.
	\newblock Network analysis reveals strongly localized impacts of el ni{\~n}o.
	\newblock {\em Proceedings of the National Academy of Sciences},
	114(29):7543--7548, 2017.
	
	\bibitem{donges2009complex}
	Jonathan~F Donges et~al.
	\newblock Complex networks in climate dynamics.
	\newblock {\em The European Physical Journal Special Topics}, 174(1):157--179,
	2009.
	
	\bibitem{kovacs2019network}
	Istv{\'a}n~A Kov{\'a}cs et~al.
	\newblock Network-based prediction of protein interactions.
	\newblock {\em Nature communications}, 10(1):1--8, 2019.
	
	\bibitem{de2015structural}
	Manlio De~Domenico et~al.
	\newblock Structural reducibility of multilayer networks.
	\newblock {\em Nature communications}, 6(1):1--9, 2015.
	
	\bibitem{stauffer1999self}
	Dietrich Stauffer and Didier Sornette.
	\newblock Self-organized percolation model for stock market fluctuations.
	\newblock {\em Physica A: Statistical Mechanics and its Applications},
	271(3-4):496--506, 1999.
	
	\bibitem{onnela2003dynamics}
	J-P Onnela, Anirban Chakraborti, Kimmo Kaski, Janos Kertesz, and Antti Kanto.
	\newblock Dynamics of market correlations: Taxonomy and portfolio analysis.
	\newblock {\em Physical Review E}, 68(5):056110, 2003.
	
	\bibitem{stauffer2000sharp}
	D~Stauffer and N~Jan.
	\newblock Sharp peaks in the percolation model for stock markets.
	\newblock {\em Physica A: Statistical Mechanics and its Applications},
	277(1-2):215--219, 2000.
	
	\bibitem{dorogovtsev2008critical}
	Sergey~N Dorogovtsev, Alexander~V Goltsev, and Jos{\'e}~FF Mendes.
	\newblock Critical phenomena in complex networks.
	\newblock {\em Reviews of Modern Physics}, 80(4):1275, 2008.
	
	\bibitem{stauffer2018introduction}
	Dietrich Stauffer and Ammon Aharony.
	\newblock {\em Introduction to percolation theory}.
	\newblock CRC press, 2018.
	
	\bibitem{bunde2012fractals}
	Armin Bunde and Shlomo Havlin.
	\newblock {\em Fractals and disordered systems}.
	\newblock Springer Science \& Business Media, 2012.
	
	\bibitem{pandey1983confirmation}
	RB~Pandey and D~Stauffer.
	\newblock Confirmation of dynamical scaling at the percolation threshold.
	\newblock {\em Physical Review Letters}, 51(7):527, 1983.
	
	\bibitem{kirkpatrick1971classical}
	Scott Kirkpatrick.
	\newblock Classical transport in disordered media: scaling and effective-medium
	theories.
	\newblock {\em Physical Review Letters}, 27(25):1722, 1971.
	
	\bibitem{derrida1983transfer}
	B~Derrida, D~Stauffer, et~al.
	\newblock Transfer matrix calculation of conductivity in three-dimensional
	random resistor networks at percolation threshold.
	\newblock {\em Journal de Physique Lettres}, 44(17):701--706, 1983.
	
	\bibitem{kirkpatrick1973percolation}
	Scott Kirkpatrick.
	\newblock Percolation and conduction.
	\newblock {\em Reviews of modern physics}, 45(4):574, 1973.
	
	\bibitem{bianconi2018multilayer}
	Ginestra Bianconi.
	\newblock {\em Multilayer networks: structure and function}.
	\newblock Oxford university press, 2018.
	
	\bibitem{de2013mathematical}
	Manlio De~Domenico, Albert Sol{\'e}-Ribalta, Emanuele Cozzo, Mikko Kivel{\"a},
	Yamir Moreno, Mason~A Porter, Sergio G{\'o}mez, and Alex Arenas.
	\newblock Mathematical formulation of multilayer networks.
	\newblock {\em Physical Review X}, 3(4):041022, 2013.
	
	\bibitem{buldyrev2010catastrophic}
	Sergey~V Buldyrev et~al.
	\newblock Catastrophic cascade of failures in interdependent networks.
	\newblock {\em Nature}, 464(7291):1025, 2010.
	
	\bibitem{stippinger2014enhancing}
	Marcell Stippinger and J{\'a}nos Kert{\'e}sz.
	\newblock Enhancing resilience of interdependent networks by healing.
	\newblock {\em Physica A: Statistical Mechanics and its Applications},
	416:481--487, 2014.
	
	\bibitem{gao2012interdependentnetworks}
	Jianxi Gao et~al.
	\newblock Networks formed from interdependent networks.
	\newblock {\em Nature Physics}, 8(1):40, 2012.
	
	\bibitem{baxter2012avalanche}
	GJ~Baxter et~al.
	\newblock Avalanche collapse of interdependent networks.
	\newblock {\em Physical Review Letters}, 109(24):248701, 2012.
	
	\bibitem{gross2020interconnections}
	Bnaya Gross et~al.
	\newblock Interconnections between networks acting like an external field in a
	first-order percolation transition.
	\newblock {\em Physical Review E}, 101(2):022316, 2020.
	
	\bibitem{radicchi2015percolation}
	Filippo Radicchi.
	\newblock Percolation in real interdependent networks.
	\newblock {\em Nature Physics}, 11(7):597--602, 2015.
	
	\bibitem{danziger2016effect}
	Michael~M Danziger et~al.
	\newblock The effect of spatiality on multiplex networks.
	\newblock {\em EPL (Europhysics Letters)}, 115(3):36002, 2016.
	
	\bibitem{halu2014emergence}
	Arda Halu, Satyam Mukherjee, and Ginestra Bianconi.
	\newblock Emergence of overlap in ensembles of spatial multiplexes and
	statistical mechanics of spatial interacting network ensembles.
	\newblock {\em Physical Review E}, 89(1):012806, 2014.
	
	\bibitem{bullmore2012economy}
	Ed~Bullmore and Olaf Sporns.
	\newblock The economy of brain network organization.
	\newblock {\em Nature Reviews Neuroscience}, 13(5):336--349, 2012.
	
	\bibitem{markov2014weighted}
	Nikola~T Markov, MM~Ercsey-Ravasz, AR~Ribeiro~Gomes, Camille Lamy, Loic Magrou,
	Julien Vezoli, P~Misery, A~Falchier, R~Quilodran, MA~Gariel, et~al.
	\newblock A weighted and directed interareal connectivity matrix for macaque
	cerebral cortex.
	\newblock {\em Cerebral cortex}, 24(1):17--36, 2014.
	
	\bibitem{ercsey2013predictive}
	M{\'a}ria Ercsey-Ravasz, Nikola~T Markov, Camille Lamy, David~C Van~Essen,
	Kenneth Knoblauch, Zolt{\'a}n Toroczkai, and Henry Kennedy.
	\newblock A predictive network model of cerebral cortical connectivity based on
	a distance rule.
	\newblock {\em Neuron}, 80(1):184--197, 2013.
	
	\bibitem{horvat2016spatial}
	Szabolcs Horv{\'a}t, R{\u{a}}zvan G{\u{a}}m{\u{a}}nuț, M{\'a}ria
	Ercsey-Ravasz, Lo{\"\i}c Magrou, Bianca G{\u{a}}m{\u{a}}nuț, David~C
	Van~Essen, Andreas Burkhalter, Kenneth Knoblauch, Zolt{\'a}n Toroczkai, and
	Henry Kennedy.
	\newblock Spatial embedding and wiring cost constrain the functional layout of
	the cortical network of rodents and primates.
	\newblock {\em PLoS biology}, 14(7):e1002512, 2016.
	
	\bibitem{gross2017multi}
	Bnaya Gross et~al.
	\newblock Multi-universality and localized attacks in spatially embedded
	networks.
	\newblock In {\em Proceedings of the Asia-Pacific Econophysics Conference
		2016—Big Data Analysis and Modeling toward Super Smart
		Society—(APEC-SSS2016)}, page 011002, 2017.
	
	\bibitem{vaknin2017spreading}
	Dana Vaknin, Michael~M Danziger, and Shlomo Havlin.
	\newblock Spreading of localized attacks in spatial multiplex networks.
	\newblock {\em New Journal of Physics}, 19(7):073037, 2017.
	
	\bibitem{bonamassa2019critical}
	Ivan Bonamassa et~al.
	\newblock Critical stretching of mean-field regimes in spatial networks.
	\newblock {\em Physical Review Letters}, 123(8):088301, 2019.
	
	\bibitem{cellai2011tricritical}
	Davide Cellai et~al.
	\newblock Tricritical point in heterogeneous k-core percolation.
	\newblock {\em Physical Review Letters}, 107(17):175703, 2011.
	
	\bibitem{panduranga2017generalized}
	Nagendra~K Panduranga et~al.
	\newblock Generalized model for k-core percolation and interdependent networks.
	\newblock {\em Physical Review E}, 96(3):032317, 2017.
	
	\bibitem{baxter2011heterogeneous}
	Gareth~J Baxter et~al.
	\newblock Heterogeneous k-core versus bootstrap percolation on complex
	networks.
	\newblock {\em Physical Review E}, 83(5):051134, 2011.
	
	\bibitem{danziger2015interdependent}
	Michael~M Danziger, Amir Bashan, and Shlomo Havlin.
	\newblock Interdependent resistor networks with process-based dependency.
	\newblock {\em New Journal of Physics}, 17(4):043046, 2015.
	
	\bibitem{morris2012transport}
	Richard~G Morris and Marc Barthelemy.
	\newblock Transport on coupled spatial networks.
	\newblock {\em Physical Review Letters}, 109(12):128703, 2012.
	
	\bibitem{morris2013interdependent}
	Richard~G Morris and Marc Barthelemy.
	\newblock Interdependent networks: the fragility of control.
	\newblock {\em Scientific reports}, 3:2764, 2013.
	
	\bibitem{morris2014spatial}
	Richard~G Morris and Marc Barthelemy.
	\newblock Spatial effects: Transport on interdependent networks.
	\newblock In {\em Networks of networks: the last frontier of complexity}, pages
	145--161. Springer, 2014.
	
	\bibitem{danziger2020faster}
	Michael~M Danziger, Bnaya Gross, and Sergey~V Buldyrev.
	\newblock Faster calculation of the percolation correlation length on spatial
	networks.
	\newblock {\em Physical Review E}, 101(1):013306, 2020.
	
	\bibitem{grassberger1999conductivity}
	Peter Grassberger.
	\newblock Conductivity exponent and backbone dimension in 2-d percolation.
	\newblock {\em Physica A: Statistical Mechanics and its Applications},
	262(3-4):251--263, 1999.
	
	\bibitem{unger1984nucleation}
	Chris Unger and W~Klein.
	\newblock Nucleation theory near the classical spinodal.
	\newblock {\em Physical Review B}, 29(5):2698, 1984.
	
	\bibitem{congilio1980clusters}
	A~Congilio and W~Klein.
	\newblock Clusters and ising critical droplets: A renormalization group
	approach.
	\newblock {\em Journal of Physics A}, 13:2775, 1980.
	
	\bibitem{heermann1983nucleation}
	Dieter~W Heermann and W~Klein.
	\newblock Nucleation and growth of nonclassical droplets.
	\newblock {\em Physical Review Letters}, 50(14):1062, 1983.
	
	\bibitem{li2012cascading}
	Wei Li, Amir Bashan, Sergey~V Buldyrev, H~Eugene Stanley, and Shlomo Havlin.
	\newblock Cascading failures in interdependent lattice networks: The critical
	role of the length of dependency links.
	\newblock {\em Physical Review Letters}, 108(22):228702, 2012.
	
	\bibitem{dorogovtsev2006k}
	Sergey~N Dorogovtsev, Alexander~V Goltsev, and Jose Ferreira~F Mendes.
	\newblock K-core organization of complex networks.
	\newblock {\em Physical Review Letters}, 96(4):040601, 2006.
	
	\bibitem{radicchi2017redundant}
	Filippo Radicchi and Ginestra Bianconi.
	\newblock Redundant interdependencies boost the robustness of multiplex
	networks.
	\newblock {\em Physical Review X}, 7(1):011013, 2017.
	
	\bibitem{newman2001random}
	Mark~EJ Newman, Steven~H Strogatz, and Duncan~J Watts.
	\newblock Random graphs with arbitrary degree distributions and their
	applications.
	\newblock {\em Physical Review E}, 64(2):026118, 2001.
	
	\bibitem{newman2008bicomponents}
	MEJ Newman and Gourab Ghoshal.
	\newblock Bicomponents and the robustness of networks to failure.
	\newblock {\em Physical Review Letters}, 100(13):138701, 2008.
	
\end{thebibliography}

\end{document}